\documentclass[conference]{IEEEtran}
\IEEEoverridecommandlockouts

\usepackage{cite}
\usepackage{amsmath,amssymb,amsfonts}
\usepackage{algorithmic}
\usepackage{graphicx}
\usepackage{subcaption}
\usepackage{textcomp}
\usepackage{xcolor}
\usepackage{booktabs}
\usepackage{hyperref}
\usepackage{caption}

\def\BibTeX{{\rm B\kern-.05em{\sc i\kern-.025em b}\kern-.08em
    T\kern-.1667em\lower.7ex\hbox{E}\kern-.125emX}}

\title{TransPathNet: A Novel Two-Stage Framework for Indoor Radio Map Prediction}
\newcommand{\EEE}{\textsuperscript{1}} 
\newcommand{\TM}{\textsuperscript{2}} 

\author{
\IEEEauthorblockN{Xin Li\EEE, Ran Liu\EEE, Saihua Xu\TM, Sirajudeen Gulam Razul\TM, Chau Yuen\EEE}
\IEEEauthorblockA{\EEE\textit{School of Electrical and Electronic Engineering, Nanyang Technological University}, Singapore \\
\TM\textit{Temasek Laboratories @ NTU, Nanyang Technological University}, Singapore \\
xin019@e.ntu.edu.sg \quad  \{ran.liu, shxu, esirajudeen, chau.yuen\}@ntu.edu.sg}
\thanks{
This research is partially supported by the National Research Foundation, Singapore and Infocomm Media Development Authority under its Future Communications Research \& Development Programme FCP-SUTD-TG-2022-010.
}
}

\begin{document}
%
\maketitle
\begin{abstract}

Accurate indoor pathloss prediction is crucial for optimizing wireless communication in indoor settings, where diverse materials and complex electromagnetic interactions pose significant modeling challenges.
This paper introduces TransPathNet, a novel two-stage deep learning framework that leverages transformer-based feature extraction and multiscale convolutional attention decoding to generate high-precision indoor radio pathloss maps. 
TransPathNet demonstrates state-of-the-art performance in the ICASSP 2025 Indoor Pathloss Radio Map Prediction Challenge, achieving an overall Root Mean Squared Error (RMSE) of 10.397~dB on the challenge full test set and 9.73~dB on the challenge Kaggle test set, showing excellent generalization capabilities across different indoor geometries, frequencies, and antenna patterns.
Our project page, including the associated code, is available at \url{https://lixin.ai/TransPathNet/}.
\end{abstract}

%

\section{Introduction}
\label{sec:intro}

The growing demand for continuous wireless connectivity in complex indoor environments, such as multi-story buildings, large office complexes, and busy commercial centers, has made accurate pathloss prediction increasingly critical~\cite{liu2019collaborative,liu2023exploiting}. Traditional propagation models face significant challenges in balancing computational efficiency with accuracy for effective network planning, often requiring extensive computational resources or suffering from limited prediction accuracy~\cite{sarkar2003survey}.

Recent breakthroughs in deep learning have consequently brought new pathloss prediction models that can successfully overcome several limitations of conventional methods. Although current developments like RadioUNet~\cite{levie2021radiounet} have approached outdoor pathloss prediction and present better performance, indoor pathloss prediction remains difficult due to the wide variety of building materials and dynamic layouts that produce complex electromagnetic interactions.



To address these challenges, the ICASSP 2025 First Indoor Pathloss Radio Map Prediction Challenge was established, using a dataset derived from ray tracing simulations to evaluate and enhance the adaptability of deep learning models operating under different indoor geometries, frequencies, and antenna patterns~\cite{rmapChallenge2025}.

This paper introduces \textbf{TransPathNet}, an encoder-decoder framework designed for high-precision indoor pathloss prediction. TransPathNet combines transformer-based feature extraction with multiscale convolutional attention decoding, employing a coarse-to-fine two-stage strategy to enhance prediction accuracy. Our model achieves an overall Root Mean Squared Error (RMSE) of 10.397~dB and 9.73~dB across three weighted tasks in the test set and Kaggle test set of the ICASSP 2025 Indoor Pathloss Radio Map Prediction Challenge.

\begin{figure*}[htbp]
  \centering
  \includegraphics[width=0.72\textwidth]{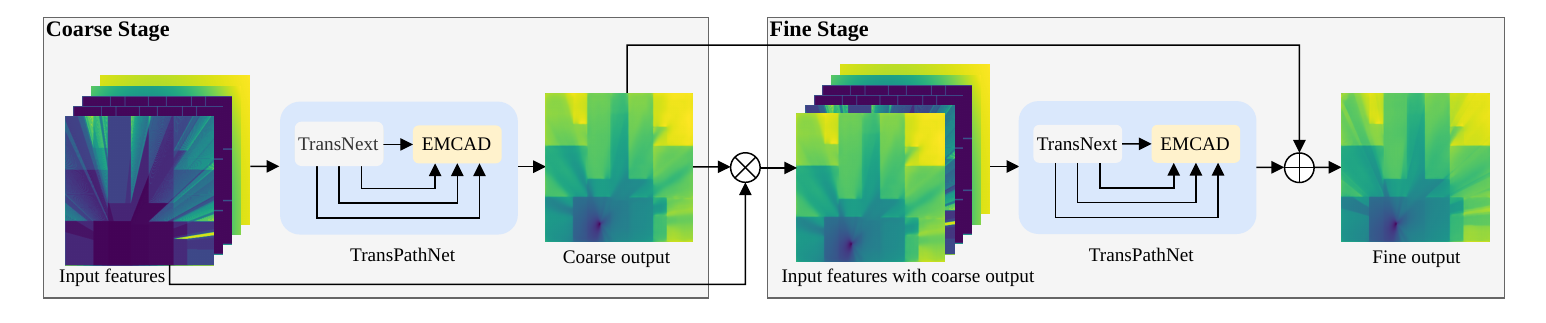}
    \caption{
    \textbf{Overview of TransPathNet training process.} The framework employs a two-stage architecture: a coarse stage and a fine stage. TransPathNet, utilized in both stages, integrates the TransNeXt and EMCAD to produce robust pathloss prediction.}
     
  \label{fig:system-overview}
\end{figure*}

\section{System Architecture}
\label{sec:system}

As shown in Fig.~\ref{fig:system-overview}, the proposed system employs a two-stage framework consisting of a coarse stage and a fine stage. Both stages use TransPathNet for pathloss prediction.

\subsection{TransPathNet}


TransPathNet follows a U-Net-like architecture with a transformer-based encoder and multi-scale convolutional attention-based decoder.
In particular, the encoder is based on TransNeXt~\cite{shi2023transnext}, a state-of-the-art backbone extracting hierarchical features from complex environmental data toward improving robustness and accuracy of pathloss predictions.
We incorporate the Efficient Multiscale Convolutional Attention Decoder (EMCAD)~\cite{rahman2024emcad}, which refines and reconstructs path loss maps at multiple scales through an attention mechanism.


\subsection{Two-Stage Training Strategy}

Our system employs a two-stage coarse-to-fine training strategy to achieve high-precision prediction results. 
First, the coarse model generates a rough approximation of the pathloss map. 
Then, the coarse result is concatenated with the input features and fed into the second refined model, which focuses on the residual details of the target and the coarse result. 

\subsection{Input Features Enhancement}
\label{subsec:features}

To capture the complexity of indoor propagation, TransPathNet extends the default three-channel inputs (reflectance, transmittance, distance) with:
Free Space PathLoss (FSPL): Precomputed free-space pathloss estimate.
Transmission Ray Encoding: Precomputed direct transmissions that highlight multi-path effects.
Antenna Embeddings: Encodes both the antenna’s pattern and angle information.
Spatial-Frequency Embeddings: Combines positional encoding and frequency embedding.
These additional channels collectively aid the model in generalizing to diverse indoor layouts, materials, and operating conditions.

\subsection{Post-Processing}

To enhance robustness, during evaluation we apply rotations and flips to the input features and perform ensembling by averaging the predictions to obtain the final output.

\begin{table}[!htbp]
\centering
\caption{Ablation study results of pathloss prediction performance for different model configurations.}
\resizebox{0.45\textwidth}{!}{
\centering
\begin{tabular}{l||c c  ||c c}
\toprule
\textit{\textbf{Case}} & \textit{\textbf{Two-Stage}} & \textit{\textbf{Post-Process}} & \textit{\textbf{RMSE(dB): Kaggle}} $\downarrow$  &\textit{\textbf{RMSE(dB): full}} $\downarrow$\\ 
\midrule
Coarse only& $\times$           & $\times$           & 9.93&10.327\\ 
+ Two-Stage Training& $\checkmark$       & $\times$           & 9.75&10.430\\ 
Full pipeline& $\checkmark$       & $\checkmark$       & 9.73&10.397 \\ 
\bottomrule
\end{tabular}
}
\label{tab:results}
\end{table}

\section{Experiments}
\label{sec:experiment}

\subsection{Implementation Details}

The model is implemented in PyTorch and trained using the Adam optimizer with an initial learning rate of $10^{-4}$, halved at 50\% and 75\% of training progress. The input features are resized to $384 \times 384$ across all training and evaluation set. The original input features are randomly flipped and rotated to improve generalizability. The Mean Squared Error (MSE) loss was chosen in training. Training was conducted on an NVIDIA RTX 4090 GPU with a batch size of 4 for 30 epochs.

\begin{figure}[!htbp]
\centering
  \includegraphics[width=0.45\textwidth]{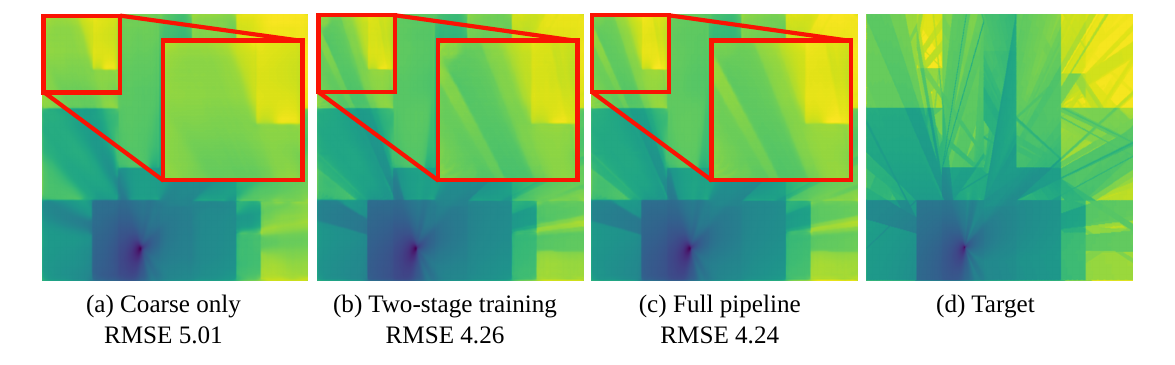}
  \caption{Visual comparison of pathloss predictions across different stages of our pipeline for a particular input.}
  \label{fig:comparision}
\end{figure}

\subsection{Results and Analysis}

RMSE in dB is the main metric for evaluating the model. The test dataset is divided into three tasks, each of which aims to evaluate the adaptability of the model to new (1) geometric environments, (2) frequencies, and (3) antenna patterns. The weights of three tasks are 30\%, 30\%, and 40\%, respectively.

We have conducted an ablation study to evaluate the impact of our model enhancements. The results are enumerated in TABLE~\ref{tab:results}. 
On the Kaggle subset, the baseline (coarse only) model achieves an RMSE of 9.93~dB, which improves to 9.75~dB with our two-stage training and 9.73~dB with the full post-processing pipeline. 
However, these gains do not generalize to the full test set, where the coarse only model performs best at 10.327~dB, which may be due to excessive reference to Kaggle data results during the competition, resulting in overfitting on the Kaggle data.
The average inference time is about 43.8\,ms per sample on the RTX 4090.
Fig.~\ref{fig:comparision} illustrates the progressive improvements in prediction quality across our pipeline stages, from coarse prediction to post-processed results.

\section{Conclusion}
\label{sec:conclusion}
This paper presents TransPathNet, an advanced deep learning framework for indoor pathloss prediction that combines transformer-based feature extraction with multi-scale convolutional attention decoding.
Our model achieves state-of-the-art performance in the ICASSP 2025 Indoor Pathloss Radio Map Prediction Challenge, demonstrating robust generalization across different geometries, frequencies, and antenna patterns.
However, it is still difficult to predict high quality pathloss caused by reflections. Future work will focus on developing network designs to improve the accuracy of these predictions.




\bibliographystyle{IEEEbib}
\bibliography{refs}

\begin{thebibliography}{1}

\bibitem{liu2019collaborative}
Ran Liu, Sumudu~Hasala Marakkalage, Madhushanka Padmal, Thiruketheeswaran
  Shaganan, Chau Yuen, Yong~Liang Guan, and U-Xuan Tan,
\newblock ``{Collaborative SLAM based on WiFi fingerprint similarity and motion
  information},''
\newblock {\em IEEE Internet Things J.}, vol. 7, no. 3, pp. 1826--1840, 2019.

\bibitem{liu2023exploiting}
Ran Liu, Billy Pik~Lik Lau, Khairuldanial Ismail, Achala Chathuranga, Chau
  Yuen, Simon~X Yang, Yong~Liang Guan, Shiwen Mao, and U-Xuan Tan,
\newblock ``{Exploiting Radio Fingerprints for Simultaneous Localization and
  Mapping},''
\newblock {\em IEEE Pervasive Comput.}, vol. 22, no. 3, pp. 38--46, 2023.

\bibitem{sarkar2003survey}
Tapan~K Sarkar, Zhong Ji, Kyungjung Kim, Abdellatif Medouri, and Magdalena
  Salazar-Palma,
\newblock ``{A survey of various propagation models for mobile
  communication},''
\newblock {\em IEEE Antennas Propag. Mag.}, vol. 45, no. 3, pp. 51--82, 2003.

\bibitem{levie2021radiounet}
Ron Levie, {\c{C}}a{\u{g}}kan Yapar, Gitta Kutyniok, and Giuseppe Caire,
\newblock ``{RadioUNet: Fast radio map estimation with convolutional neural
  networks},''
\newblock {\em IEEE Trans. Wireless Commun.}, vol. 20, no. 6, pp. 4001--4015,
  2021.

\bibitem{rmapChallenge2025}
Stefanos Bakirtzis, Cagkan Yapar, Kehai Qui, Ian Wassell, and Jie Zhang,
\newblock ``{The First Indoor Pathloss Radio Map Prediction Challenge},''
\newblock in {\em Proc. IEEE International Conference on Acoustics, Speech, and
  Signal Processing (ICASSP)}, April 2025.

\bibitem{shi2023transnext}
Dai Shi,
\newblock ``{TransNeXt: Robust Foveal Visual Perception for Vision
  Transformers},''
\newblock in {\em Proc. IEEE/CVF Conf. Comput. Vis. Pattern Recognit. (CVPR)},
  June 2024, pp. 17773--17783.

\bibitem{rahman2024emcad}
Md~Mostafijur Rahman, Mustafa Munir, and Radu Marculescu,
\newblock ``{EMCAD: Efficient multi-scale convolutional attention decoding for
  medical image segmentation},''
\newblock in {\em Proc. IEEE/CVF Conf. Comput. Vis. Pattern Recognit. (CVPR)},
  2024, pp. 11769--11779.

\end{thebibliography}

\end{document}